\documentclass[12pt]{article}
\pdfoutput=1
\usepackage{amsmath,amssymb,amsfonts,amsthm,bm,bbm,cancel,wasysym}
\usepackage{epsfig,graphics,graphicx,epstopdf,caption,subcaption}
\usepackage[numbers,sort&compress]{natbib}
\graphicspath{{Charts/}}
\usepackage{array,booktabs,colortbl,colordvi,multirow}
\usepackage{colordvi,color,xcolor}
\usepackage{hyperref}
\usepackage{rotating}
\usepackage{comment}
\usepackage{feynmf}

\begin{document}
\begin{center}

{\Large {\bf  Resolving Muon g-2 Anomaly With Partial Compositeness }}\\

\vspace*{0.75cm}

{Shuai Xu$^1$ and Sibo Zheng$^2$}

\vspace{0.5cm}
{$^1$School of Physics and Telecommunications Engineering, Zhoukou Normal University, Henan 466001, China\\
$^2$Department of Physics, Chongqing University, Chongqing 401331, China
}
\end{center}
\vspace{.5cm}

\begin{abstract}
We consider the scenario of composite Higgs with partial compositeness to address the muon g-2 anomaly. 
We show that this anomaly is resolved by one-loop correction due to composite muon partners with mass scale of order TeVs and large Yukawa coupling to composite Higgs, which is different from interpretations of vectorlike lepton models.
We present parameter space by imposing indirect constraints from precise measurements on Higgs, Z and oblique electroweak parameters.
We analyze direct constraints in light of both Drell-Yan and Higgs-associated productions of the composite muon partners at high-luminosity LHC.
It turns out that the parameter regions with the lighter composite muon mass below $\sim 325$ GeV can be excluded at 2$\sigma$ order by the Drell-Yan processes.
\end{abstract}

\renewcommand{\thefootnote}{\arabic{footnote}}
\setcounter{footnote}{0}
\thispagestyle{empty}
\vfill
\newpage
\setcounter{page}{1}

\tableofcontents
\section{Introduction}
The muon anomalous magnetic moment $a_{\mu}$ is an effective operator of dimension five
\begin{eqnarray}{\label{amu}}
\mathcal{L}\supset \frac{e}{4m_{\mu}}a_{\mu} \bar{u}\sigma_{\rho\lambda} u F^{\rho\lambda},
\end{eqnarray}
where $e$ is the electric charge, $m_{\mu}$ is the muon mass, $\sigma^{\rho\lambda}=\frac{i}{2}[\gamma^{\rho},\gamma^{\lambda}]$, and $F_{\rho\lambda}$ is the electromagnetic field strength.
Measurement of this observable provides a direct check on the completeness of a renormalizable theory such as Standard Model (SM).
This makes $a_{\mu}$ one of those precisely calculated in the SM, see \cite{Aoyama:2020ynm, 0902.3360} for reviews.

A comparison between the theoretical prediction of the SM and precise measurements on $a_{\mu}$ enables us to uncover imprints of new physics.
A so-called muon anomaly has been reported by the BNL \cite{Muong-2:2006rrc} experiment, 
which shows a $3.7\sigma$ deviation of $a_{\mu}$ from the SM reference value.\footnote{The SM theoretical value of $a_{\mu}$  is subject to an important uncertainty arising from QCD hadron effects \cite{Borsanyi:2020mff}.}
Recently,  this deviation is further strengthened by the latest FNAL \cite{Muong-2:2021ojo} experiment with a larger significance of $4.2\sigma$. Combining the BNL and FNAL data gives
\begin{eqnarray}{\label{deltaamu}}
\delta a^{\rm{exp}}_{\mu}= (2.51\pm 0.59) \times 10^{-9}.
\end{eqnarray}

To explain this anomaly, various models of new physics have been proposed in the literature.
Since $\delta a^{\rm{exp}}_{\mu}$ in Eq.(\ref{deltaamu}) is of order one-loop electroweak contribution to $a_{\mu}$, 
it is natural to consider new electroweak degrees of freedom such as vectorlike leptons \cite{Kannike:2011ng, Dermisek:2013gta, Freitas:2014pua,Poh:2017tfo,Dermisek:2021ajd,DeJesus:2020yqx, Frank:2020smf,Arkani-Hamed:2021xlp} among others, see \cite{Athron:2021iuf} for a comprehensive review on model buildings.
In this work, we consider an alternative scenario of composite Higgs with partial compositeness,
where composite muon partners play the role of the vectorlike leptons.
Previous work, 
which advocated interpreting the muon anomaly with partial compositeness, 
can be found in \cite{Redi:2013pga,Antipin:2014mda,Doff:2015nru, Megias:2017dzd, Cacciapaglia:2021gff}.
In this paper, we will consider this scenario by a systematic analysis, based on a complete effective field theory description of this scenario.
As in the aforementioned vectorlike lepton models,
the desired contribution to $a_{\mu}$ due to the composite lepton partners
is expected to be constrained by their effects on the other precise electroweak observables,
either through gauge and/or Yukawa interactions.

The rest of this paper is organized as follows.
In Sec.\ref{model}, we introduce the effective field theory description of our scenario,
with key features different from those of vectorlike leptons emphasized.
Sec.\ref{anomaly} addresses the new contributions to $a_{\mu}$ due to the composite quark and composite lepton partners, 
which are two- and one-loop effect respectively. 
We will show that the significance of the observed anomaly can be naturally reduced to 2$\sigma$ order by the composite muon partners with mass scale of order $\sim$ TeVs and strong coupling constant $g_{*}$ of order unity.
In Sec.\ref{constraint}, we firstly consider indirect constraints from precision measurements on Higgs, $Z$ and oblique electroweak parameters, then discuss the prospect of searching for signals in the parameter space 
in terms of both Drell-Yan and Higgs-associated productions of the composite muon partners at the high-luminosity LHC. 
We present the main numerical results in Sec.\ref{results}.
Finally, we conclude in Sec.\ref{con}.

\section{Partial Compositeness}
\label{model}
In the framework of composite Higgs with partial compositeness, 
as firstly proposed by refs.\cite{Dugan:1984hq, Kaplan:1991dc},
the general effective Lagrangian for partially composite fermion degrees of freedom based on coset $\mathcal{G}/\mathcal{H}$ reads as
\begin{eqnarray}{\label{Lag}}
\mathcal{L}&=&\mathcal{L}_{\rm{e}}+\mathcal{L}_{\rm{s}}+\mathcal{L}_{\rm{m}}
\end{eqnarray}
where the elementary sector, the strong sector, and the mixing effects are respectively,
\begin{eqnarray}{\label{Lag1}}
\mathcal{L}_{\rm{e}}&=&i\bar{q}_{L}\gamma^{\mu}D_{\mu}q_{L}+i\bar{u}_{R}\gamma^{\mu}D_{\mu}u_{R}+i\bar{d}_{R}\gamma^{\mu}D_{\mu}d_{R}+i\bar{\ell}\gamma^{\mu}D_{\mu}\ell+i\bar{e}_{R}\gamma^{\mu}D_{\mu}e_{R},\nonumber\\
\mathcal{L}_{\rm{s}}&=& \frac{f^{2}}{4}D_{\mu}U (D^{\mu}U)^{\dag}+\bar{\Psi}\left(i\gamma^{\mu}D_{\mu}-m_{\Psi}\right)\Psi\nonumber\\
&-&\left[g_{*}\left(Q_{L}\bar{U}_{R}H+Q_{L}\bar{D}_{R}\tilde{H}+L_{L}\bar{E}_{R}\tilde{H}\right)+g_{*}\left(Q_{R}\bar{U}_{L}H+Q_{R}\bar{D}_{L}\tilde{H}+L_{R}\bar{E}_{L}\tilde{H}\right)+h.c.\right]\nonumber\\
&+& \rm{higher-dimensional~terms},\nonumber\\
\mathcal{L}_{\rm{m}}&=&-\sin\theta_{\ell_{L}}m_{L}\bar{\ell}_{L}L_{R}-\sin\theta_{\ell_{R}}m_{E}\bar{E}_{L}e_{R}+\cdots+\rm{h.c}.
\end{eqnarray}
Here, $q_{L}$ and $\ell$ are the SM quark and lepton doublets respectively whereas $u_{R}$, $d_{R}$ and $e_{R}$ are the $SU(2)_{L}$ singlets;
with respect to these SM states $Q$, $L$, $U$, $D$ and $E$ are the corresponding vectorlike composite fermions that are components of representations of $\mathcal{G}$;
$H=(H^{+}, \upsilon+ h)^{T}$ is the composite Higgs doublet, 
a pseudo Nambu-Goldstone boson described by the Goldstone matrix $U$ in Callan-Coleman-Wess-Zumino formalism \cite{Panico:2015jxa} with $\upsilon$ the weak scale and $\tilde{H}=i\sigma_{2}H^{*}$;
and $\Psi$ denotes any of the above composite fermions.
Note, besides the global group $\mathcal{G}/\mathcal{H}$ two external groups, i.e, the  QCD gauge group and $U(1)_X$ global group, are also necessary, see \cite{Panico:2015jxa} for a review.

The masses and couplings in Eq.(\ref{Lag1}) have the following physical contents,
where we will point out the key differences between this composite model and the aforementioned vectorlike models.
\begin{itemize}
\item The first term in $\mathcal{L}_{\rm{s}}$ is the kinetic term for the composite Higgs with $f$ the $\mathcal{G}$-breaking mass scale, which includes both derivative and non-derivative interactions involving the composite Higgs. While these derivative interactions can help us distinguish a composite Higgs model from vectorlike models, 
they are irrelevant for the topic of $a_{\mu}$ under consideration.
\item  The second term in $\mathcal{L}_{\rm{s}}$ refers to the kinetic term for the composite fermions with $m_{\Psi}$ their vectorlike masses,
where the gauge covariant derivative relies on explicit representations of the composite states.
Here, we simply adopt $D_{\mu}=\partial_{\mu}-iYg'B_{\mu}-ig W_{\mu}$,\footnote{We will reminder the reader when arguments related to this point are made.}
where $B_{\mu}$ is the $U(1)_Y$ gauge field with gauge coupling  $g'$  and hypercharge $Y$,
and $W_{\mu}$ is the $SU(2)_L$ gauge field with gauge coupling $g$.
\item The last three terms in $\mathcal{L}_{\rm{s}}$ represent interactions between the composite Higgs and the composite fermions, which include Yukawa terms \cite{Contino:2006nn} and higher-dimensional terms such as derivative interactions \cite{Panico:2015jxa}.
The first class of Yukawa terms is responsible for reproduction of the SM Yukawa couplings after rotating to mass basis and the high-dimensional terms are instead model-dependent
with magnitudes of coefficients in these terms suppressed by power laws of $f$.
As long as $f$ is large enough their contribution to $a_{\mu}$ is subdominant compared to the Yukawa contribution.
Without losing simplicity, we will neglect these high-dimensional terms.
Note that the Yukawa interactions give rise to mixing effects between the composite doublets and singlets after electroweak symmetry breaking (EWSB). 
\item Finally, the mixing terms between the elementary and composite fermions in $\mathcal{L}_{\rm{m}}$
are essential to explicitly break $\mathcal{G}$ into $\mathcal{H}$ as characterized by the small mixing angles $\sin\theta_{\ell_{L}}$ and $\sin\theta_{\ell_{R}}$ etc, which are external to EWSB. 
Together with the mixing effects between the composite fermions in $\mathcal{L}_{s}$ induced by EWSB,
these mixing terms allow the composite muon fermions to  couple to both the left- and right-hand muon via $h$ simultaneously.
This feature is crucial to understand why the dimension of the operator in Eq.(\ref{amu}) is fixed to be five  in the composite Higgs model, 
compared to the dimension of the effective operator either the same as in some vectorlike models such as  \cite{Dermisek:2013gta}  or higher in other vectorlike models such as \cite{Arkani-Hamed:2021xlp}.
\end{itemize}
Despite the lepton compositeness being small,
the Yukawa coupling constant $g_*$ of the composite lepton fermions to the composite Higgs can be as large as $\sim \sqrt{4\pi}$, which implies that the one-loop composite muon contribution to $a_{\mu}$ can be significant even with their mass scales $m_{L}$ and $m_{E}$ bigger than $\upsilon$, as compared to light degree(s) of freedom required by the vectorlike models, among others.

Along this line both hierarchical and diagonal mixing angles in $\mathcal{L}_{m}$ in flavor space are technically appropriate to address the elementary fermion mass hierarchy and small flavor-violation effects simultaneously. 
For a quantitative analysis on the composite fermion contributions to $a_{\mu}$,
it is sufficient to make use of a set of parameters as composed of composite partner masses $m_{L}$ and $m_E$, 
the strong coupling constant $g_{*}$, and the small mixing angles $\sin\theta_{\mu_{L}}$ and $\sin\theta_{\mu_{R}}$.

\section{Muon g-2}
\label{anomaly}
In this section we discuss the contributions of new composite fermion partners to muon $a_{\mu}$.
We will show that  while the effect of composite quark partners is negligible,
the composite muon partners are able to address the anomaly.

\subsection{Composite Quark Partners}
Composite quark states contribute to $a_{\mu}$ either via
hadron effect in the low energy region or two-loop electroweak corrections. 
For the hadron effect, it goes through photon vacuum polarization \cite{0902.3360}
\begin{eqnarray}{\label{had}}
\delta a^{\rm{had}}_{\mu}\sim \int \frac{R(s)\hat{K}(s)}{s^{2}} ds, 
\end{eqnarray}
where $R(s)$ is the total cross section of electron pair annihilation with $s$ the center-of-mass energy,
and $\hat{K}$ is the weight function.
As shown in Eq.(\ref{had}), the hadron effect is proportional to $s^{-2}$ that dramatically suppresses the heavy color state contribution.
In the composite quark mass range $m_{\Psi_{C}}>m_{t}$, 
$\delta a_{\mu}(\rm{had})< 1\%\times a^{SM}_{\mu}(\rm{had})$, 
where $a^{SM}_{\mu}(\rm{had})\approx (69.0 \pm0.526) \times 10^{-9}$ \cite{0902.3360} is the SM hadron contribution.

Given the large coupling $g_*$,  the two-loop electroweak corrections to $a_{\mu}$ is dominated by a type of Barr-Zee \cite{Barr:1990vd} diagram.
In this diagram, the composite quark partners run in the inner fermion loop which gives an effective vertex of type $\Gamma_{\gamma\gamma h}$, 
and the mixing mass between the elementary and composite muon as a vertex has to be inserted in the fermion line attached to the external muon in order to flip the muon chirality. 
Compared to this diagram, 
the other two-loop Feynman diagrams as those of SM \cite{Czarnecki:1995wq,Peris:1995bb} are relatively subdominant.
Referring to the illustrative numbers that the model-independent LHC bound on the composite quark mass is larger than $\sim 1.5$ TeV \cite{ATLAS:2018ziw, CMS:2018zkf}, the model-dependent LHC bound on the composite muon mass is above $\sim 200$ GeV \cite{CMS:2019hsm, ATLAS:2015qoy}, 
and the theoretical limit on the strong coupling $g_{*}\leq \sqrt{4\pi}$,
one finds that this diagram contributes to a value of $\delta a_{\mu}$ beneath the reported value in Eq.(\ref{deltaamu}) at least by two orders of magnitude.
Adjusting the value of $g_*$ and/or $m_{\Psi_{\mu}}$ in allowed ranges does not  change the situation.

\subsection{Composite Lepton Partners}
The small mixings between the composite muon partners and the elementary muon in Eq.(\ref{Lag}) lead to the following interactions
\begin{eqnarray}{\label{leptona}}
\mathcal{L}_{\rm{int}}&=&\left[g^{W}_{NL}\bar{N}\gamma^{\mu}\mu_{L}+g^{W}_{NR}\bar{N}\gamma^{\mu}\mu_{R}\right]W^{+}_{\mu}+h.c\nonumber\\
&+&\left[g^{H}_{FL}\bar{\Psi}_{F}\mu_{L}+g^{H}_{FR}\bar{\Psi}_{F}\mu_{R}\right]h+h.c,
\end{eqnarray}
where the couplings are explicitly shown in Table.\ref{couplings}.
Here, index $F= A, B$ refer to the two mass eigenstates of the charged left- and right-hand composite muon partners with mixing angle $\beta$, see Eq.(\ref{masses}).
In Eq.(\ref{leptona}) the absence of vertex $\Psi_{F}-\mu-\gamma$ is due to the same electric charge $Q_{\Psi_{\mu}}=Q_{\mu}=-1$,
whereas that of vertex $\Psi_{F}-\mu-Z$ follows from the same weak isospin we have adopted.\footnote{
This vertex can be present in multi-site realizations of the scenario, 
where the gauge covariant derivative takes a form $D_{\mu}=\partial_{\mu}-ig'Y B_{\mu}-ig_{*}\rho_{\mu}$, 
with $\rho_{\mu}$ heavy composite gauge bosons to mix with elementary electroweak gauge bosons.
Even so, the effect of this interaction on $a_{\mu}$ is still small compared to that of the Yukawa interaction with large $g_{*}$, see below Eq.(\ref{lepton}).}

\begin{table}
\begin{center}
\begin{tabular}{cccc}
\hline\hline
 ~~~~& $A$~~~~& $B$~~~~~~& $N$\\ \hline
 $g^{W}_{NL}$~~ &  $-$~~~~ &  $-$~~~~ & $\frac{g}{\sqrt{2}}\sin\theta_{\mu_{L}}$ \\
$g^{W}_{NR}$~~ &   $-$~~~~ & $-$~~~~ &  $0$  \\
$g^{H}_{FL}$~~ &  $-g_{*}\sin\theta_{\mu_{L}}\sin\beta$ ~~~~& $g_{*}\sin\theta_{\mu_{L}}\cos\beta$ & $-$ \\
$g^{H}_{FR}$~~ &  $g_{*}\sin\theta_{\mu_{R}}\cos\beta$~~~~ & $g_{*}\sin\theta_{\mu_{R}}\sin\beta$  &  $-$ \\
\hline \hline
\end{tabular}
\caption{Couplings of the composite lepton partners to the elementary left- and right-hand muon with index $F=A,B$ in the leading order of $\sin\theta_{\mu_{L}}$ and/or $\sin\theta_{\mu_{R}}$.  See texts for details about the absences of both $\gamma$- and $Z$-relevant vertex.}
\label{couplings}
\end{center}
\end{table}

As expected, one obtains the SM muon Yukawa coupling 
\begin{eqnarray}{\label{Yukawa}}
y_{\mu}\approx g_{*}\sin\theta_{\mu_{L}} \sin\theta_{\mu_{R}},
\end{eqnarray}
after integrating out the composite muon partners $L$ and $E$.
To derive the masses $m_F$ for the eigenstates $\Psi_F$ in Eq.(\ref{leptona})
it is sufficient to only consider the mass matrix 
\begin{eqnarray}{\label{massmatrix}}
\left(\begin{array}{cc} 
\bar{L}_{L} & \bar{E}_{L} \end{array}\right)
\left(
\begin{array}{cc}
 m_{L}&  -g_{*}\upsilon \\
-g_{*}\upsilon &  m_{E} \\
\end{array}\right)
\left(\begin{array}{c}
L_{R}\\
E_{R} \\
\end{array}\right).
\end{eqnarray} 
from $\mathcal{L}_s$ and ignore the small mixing effects from $\mathcal{L}_{m}$ in Eq.(\ref{Lag}).
Diagonalizing the mass matrix in Eq.(\ref{massmatrix}) yields
\begin{eqnarray}{\label{masses}}
m_{A}&=& m_{L}\cos^{2}\beta+g_{*}\upsilon\sin2\beta+ m_{E}\sin^{2}\beta, \nonumber\\
m_{B}&=& m_{E}\cos^{2}\beta-g_{*}\upsilon\sin2\beta+ m_{L}\sin^{2}\beta,
\end{eqnarray}
where the angle $\beta$ satisfies
\begin{eqnarray}{\label{beta}}
\left(m_{L}-m_{E}\right)\sin2\beta=2g_{*}\upsilon\cos2\beta.
\end{eqnarray}

\begin{figure}
\centering
\includegraphics[width=13cm,height=3.5cm]{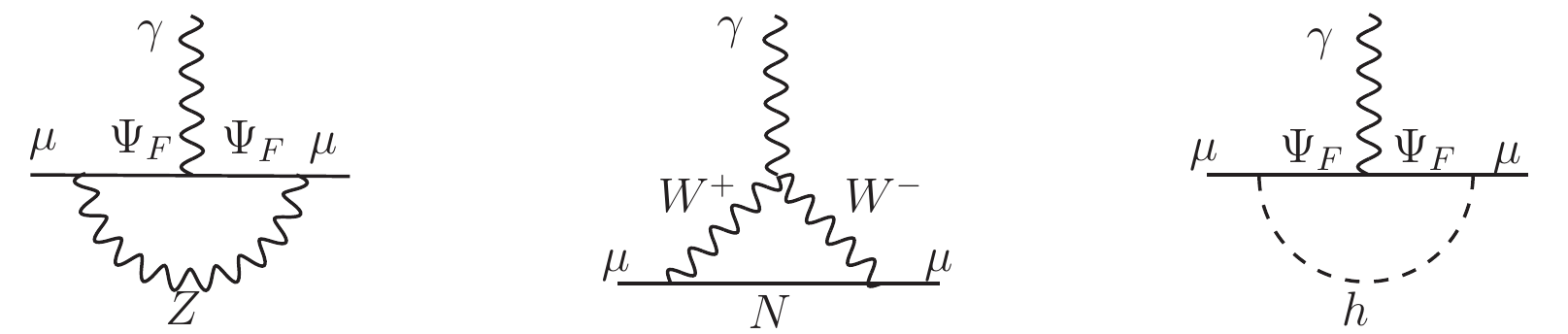}
\centering
\caption{Feynman diagrams for the new contributions to $a_{\mu}$ due to the composite states with $\Psi_F$ the composite muon states and $N$ the neutral field.}
\label{Feyn}
\end{figure}

Using Table.\ref{couplings} one obtains each contribution to $a_{\mu}$ with respect to the individual Feynman diagram in Fig.\ref{Feyn},
\begin{eqnarray}{\label{lepton}}
\delta a^{Z}_{\mu}&\approx&\sum_{F=A,B}-\frac{m^{2}_{\mu}}{16\pi^{2}m^{2}_{Z}}\left[\frac{4}{3}g^{Z}_{FL}g^{Z}_{FR}\frac{m_{F}}{m_{\mu}}\mathcal{C}\left(\frac{m^{2}_{F}}{m^{2}_{Z}}\right)\right]\nonumber\\
\delta a^{W}_{\mu}&\approx&-\frac{m^{2}_{\mu}}{16\pi^{2}m^{2}_{W}}\left[-\frac{1}{3}(g^{W}_{NL})^{2}\mathcal{M}\left(\frac{m^{2}_{N}}{m^{2}_{W}}\right)\right]\nonumber\\
\delta a^{H}_{\mu}&\approx&\sum_{F=A,B}-\frac{m^{2}_{\mu}}{16\pi^{2}m^{2}_{h}}\left[-\frac{2}{3}g^{H}_{FL}g^{H}_{FR}\frac{m_{F}}{m_{\mu}}\mathcal{F}\left(\frac{m^{2}_{F}}{m^{2}_{h}}\right)\right],
\end{eqnarray}
in the large composite mass region $m_{F}>m_{h}, m_{Z}, m_{W}$, 
where the loop functions $\mathcal{C}$, $\mathcal{M}$ and $\mathcal{F}$ can be found in ref.\cite{Athron:2021iuf}.

The correction to $a_{\mu}$ due to the composite lepton partners in Eq.(\ref{lepton}) in our model  is dominated by the $h$-exchange diagram. 
Given a fixed $g_*$ the value of $\delta a^{H}_{\mu}$ therein, which is proportional to $\sin2\beta$ as seen in Table.\ref{couplings},
maximizes at $\tan\beta=1$ that corresponds to $m_{L}\approx m_{E}\approx m_{\Psi}$.\footnote{In contrast to the degenerate masses, a large mass splitting between $m_{L}$ and $m_{E}$ leads to $\tan\beta\sim 0$ which suppresses the magnitude of $\delta a^{H}_{\mu}$.}
Under this choice $\delta a^{H}_{\mu}$ explicitly reads
\begin{eqnarray}{\label{app}}
\delta a^{H}_{\mu}\approx \frac{g_{*}y_{\mu}}{2\pi^{2}} \frac{m_{\mu}}{m_{\Psi}}f\left(\frac{g_{*}\upsilon}{m_{\Psi}}\right),
\end{eqnarray}
with the function 
\begin{eqnarray}{\label{f}}
f(x)=\frac{x}{(1-x^{2})[(1-x)^{-2}+(1+x)^{-2}]},
\end{eqnarray}
where the small mixing angles $\sin\theta_{\mu_{L,R}}$ have been replaced by $y_{\mu}$ in Eq.(\ref{Yukawa}), and the rational region of $x$ corresponds to $0<x<1$. 
Eq.(\ref{app}) shows that $\delta a^{H}_{\mu}$ are sensitive to $g_{*}$, $m_{\Psi}$ and $x$.
To resolve the muon anomaly, a large coupling $g_{*}$ seems necessary.
For instance, with $g_{*}\sim 3$ and $m_{\Psi}\sim 1$ TeV the reported muon g-2 anomaly is easily resolved.
We will present more details about the viable parameter space based on Eq.(\ref{app}) in Sec.\ref{results} after we investigate the constraints in the next section.

\section{Constraints}
\label{constraint}
\subsection{Indirect Constraints}
\label{indirect}
\subsubsection{Higgs Portal}
The strong coupling of $\Psi_F$ to the composite Higgs is the key ingredient to address the muon anomaly, 
which can be constrained by precise measurements on the Higgs couplings.
For the couplings of Higgs to SM gauge bosons we consider the one-loop modification to the coupling of $h$ to di-photon as in the effective operator $c_{\gamma}hF^{2}_{\mu\nu}$  due to the composite muon partners.
Since this operator explicitly breaks the global symmetry of $h$,
$c_{\gamma}$ is therefore controlled by the mixing terms in $\mathcal{L}_{\rm{m}}$ in Eq.(\ref{Lag}).
Inserting them into the one-loop fermion triangle diagram as vertexes gives rise to \cite{Delaunay:2013iia} 
\begin{eqnarray}{\label{Yukawaa}}
\delta c_{\gamma}\approx 
\frac{\alpha }{2\pi}\frac{g^{2}_{*}\upsilon}{m^{2}_{\Psi}}\sin^{2}\theta_{\mu_{R}}
\end{eqnarray}
in the mass region $m_{h}<m_{\Psi}< m_{*}\approx g_{*}f$.
The dependence of $\delta c_{\gamma}$ on the $\sin\theta_{\mu_{R}}$ in Eq.(\ref{Yukawaa}) implies that
the effect of the composite muon fermions on the decay $h\rightarrow \gamma\gamma$ is greatly suppressed as a result of the small $\mu$ compositeness so that it is easily beyond current LHC precision \cite{ATLAS:2019nkf, CMS:2018uag}.
Just like $h\rightarrow\gamma\gamma$ the decay mode $h\rightarrow Z\gamma$ \cite{Azatov:2013ura} imposes similar constraint, 
even though the later one is not always independently analyzed by LHC experiments.

For the coupling of Higgs to SM fermions, the precision tests on the SM Yukawa couplings provide information on the composite mass scale as 
\begin{eqnarray}{\label{Yukawad}}
\frac{\delta y_{\psi}}{y_{\psi}}\sim \frac{\upsilon^{2}}{f^{2}}.
\end{eqnarray}
Taking a conservative lower bound $f\geq 1$ TeV \cite{Xu:2020omq} inferred from current LHC data
implies that the composite mass scale $m_{*}\geq g_{*}$ TeV.

\subsubsection{$Z$ Portal}
The mixing effects between the elementary and the composite muon field modify 
the $Z-\mu-\mu$ couplings as
\begin{eqnarray}{\label{Zuu}}
g^{Z}_{\mu_{L}\mu_{L}}&\approx&\left(-\frac{1}{2}+s^{2}_{W}\right)(1+\sin^{2}\theta_{\mu_{L}}) \nonumber\\
g^{Z}_{\mu_{R}\mu_{R}}&\approx&s^{2}_{W}(1+\sin^{2}\theta_{\mu_{R}}),
\end{eqnarray}
where unity in the bracket means the SM values with $s_W$  the weak mixing angle.
Imposing the LEP constraint $\delta g^{Z}_{\mu_{L}\mu_{L}}\leq 10^{-3}$ \cite{ParticleDataGroup:2016lqr} implies 
$\sin\theta_{\mu_{L}}\leq 3.3\times 10^{-2}$, which is far above the lower bound $\sin\theta_{\mu_{L}}\geq 10^{-4}$ as seen from the SM muon Yukawa coupling in Eq.(\ref{Yukawa}).
Note that the corrections in Eq.(\ref{Zuu}) are model-dependent as they depend on
the isospins of the composite muon partners.

\subsubsection{Oblique Corrections} 
The last precise tests are oblique parameters \cite{Baak:2012kk}. 
The oblique parameter $\hat{S}$ tied to $W^{3}_{\mu}$ and $B_{\mu}$ gauge boson receives corrections from the composite vector bosons $\rho_{\mu}$ at tree level and the composite Higgs and fermions at loop level, respectively. 
The tree-level correction $\hat{S}(\rho_{\mu})\sim m^{2}_{W}/m^{2}_{*}$ implies that $m_{*}\geq 3$ TeV.
Moreover, the oblique parameter $\hat{T}$ does not receive tree-level correction due to the protection of custodial symmetry. 
$\hat{T}$ only obtains loop corrections due to the composite Higgs and fermions. 
Similar to vectorlike doublet $+$ singlet fermion contributions \cite{Lavoura:1992np,Grojean:2013qca},
the one-loop corrections to $\hat{S}$ and $\hat{T}$ due to the composite muon partner loops are 
\begin{eqnarray}{\label{ST}}
\Delta\hat{S}&\sim&\frac{g^{2}}{16\pi^{2}} \xi \ln\left(\frac{m_{*}}{m_{\Psi}}\right),\nonumber\\
\Delta\hat{T}&\sim& \frac{1}{16\pi^{2}}\xi\sin^{4}\theta_{\mu_{L}}\frac{m_{\Psi}^{2}}{f^{2}},
\end{eqnarray}
with $\xi=\upsilon^{2}/f^{2}$. 
The correction to the $\hat{T}$ parameter in Eq.(\ref{ST}) is greatly suppressed by the small factor $\sin^{4}\theta_{\mu_{L}}$ as upper bounded by the $Z$ coupling in Eq.(\ref{Zuu}).
Likewise, given $\xi\leq 0.1$ inferred from Eq.(\ref{Yukawad}) the correction to the $\hat{S}$ parameter in Eq.(\ref{ST}) can be safely neglected as well.

\subsection{Direct Constraints}
Let us now consider direct detections on the parameter space with respect to the muon anomaly.
The couplings and the small mixings of the composite muon partners $\Psi_F$ are the key factors to develop strategies of searching them at colliders,
which are obviously different from studies on composite quark partners \cite{Barcelo:2011wu, DeSimone:2012fs, Cacciapaglia:2018qep}, composite gauge bosons \cite{Thamm:2015zwa, BuarqueFranzosi:2016ooy, Liu:2018hum}, 
and composite Higgs \cite{Banerjee:2021efl}.
After produced at colliders, 
the composite muon partners decay through the modes $\Psi^{\pm}_{F}\rightarrow W^{\pm}\nu_{\mu}$, $\Psi^{\pm}_{F}\rightarrow h\mu^{\pm}$ and $\Psi^{\pm}_{F}\rightarrow W^{\pm}N$ for the heavier state $F=A$.
For a practical purpose, it is sufficient to focus on the signals of the lighter state $\Psi_B$ at the 14-TeV LHC.

For simulation of signals and relevant SM backgrounds in the following, we use FeynRules \cite{1310.1921} to generate model files prepared for MadGraph5 \cite{1405.0301} that includes Pythia 8 \cite{Sjostrand:2014zea} for parton showering and hadronization and the package Delphes 3 \cite{1307.6346} for fast detector simulation.

\begin{figure}
\centering
\includegraphics[width=8cm,height=4cm]{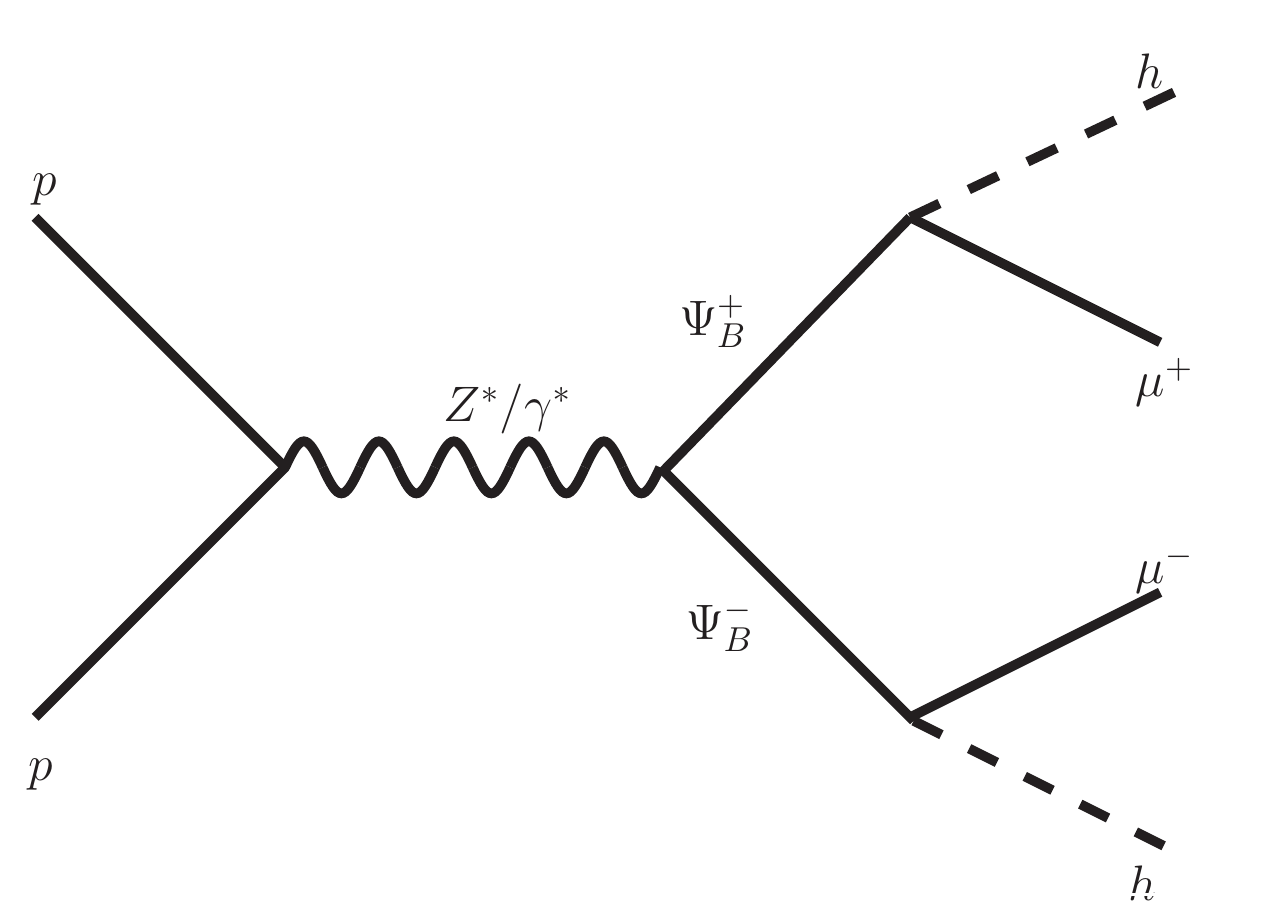}
\centering
\caption{Feynman diagram for the Drell-Yan productions of $\Psi_B$ pair at the LHC with the decay mode $\Psi^{\pm}_{B}\rightarrow h\mu^{\pm}$.}
\label{DY}
\end{figure}

\subsubsection{Gauge Boson Portal}
The $\Psi_B$ pair can be produced either through $e^{+}e^{-}\rightarrow Z^{*}/\gamma^{*}\rightarrow \Psi^{+}_{B}\Psi^{-}_{B}$ at LEP or Drell-Yan processes $pp\rightarrow Z^{*}/\gamma^{*}\rightarrow \Psi^{+}_{B}\Psi^{-}_{B}$ at LHC.
Here, do not confuse the coupling  $\bar{\Psi}_{F}-\Psi_{F}-Z$ with the coupling $\bar{\Psi}_{F}-\mu-Z$.
So far, the LEP-2 has excluded the mass range $m_{B}\leq 101.2$ GeV \cite{L3:2001xsz},
while the LHC has reported a limit $m_{B}\geq 200$ GeV based on the decay mode $\Psi^{\pm}_{B}\rightarrow Z\mu^{\pm}$ \cite{ATLAS:2015qoy} in certain vectorlike lepton models.
We can revisit this LHC bound by repeating the Drell-Yan processes with Feynman diagram shown in Fig.\ref{DY},
where the decay mode $\Psi^{\pm}_{B}\rightarrow Z\mu^{\pm}$ is replaced by our decay mode $\Psi^{\pm}_{B}\rightarrow h\mu^{\pm}$.
Followed by the decay of $h\rightarrow \bar{b}b$,
the primary SM backgrounds corresponding to this signal is given by $pp \rightarrow Z/\gamma(\ell^{+}\ell^{-})+{\rm jets}$.

\begin{figure}
\centering
\includegraphics[width=13cm,height=8cm]{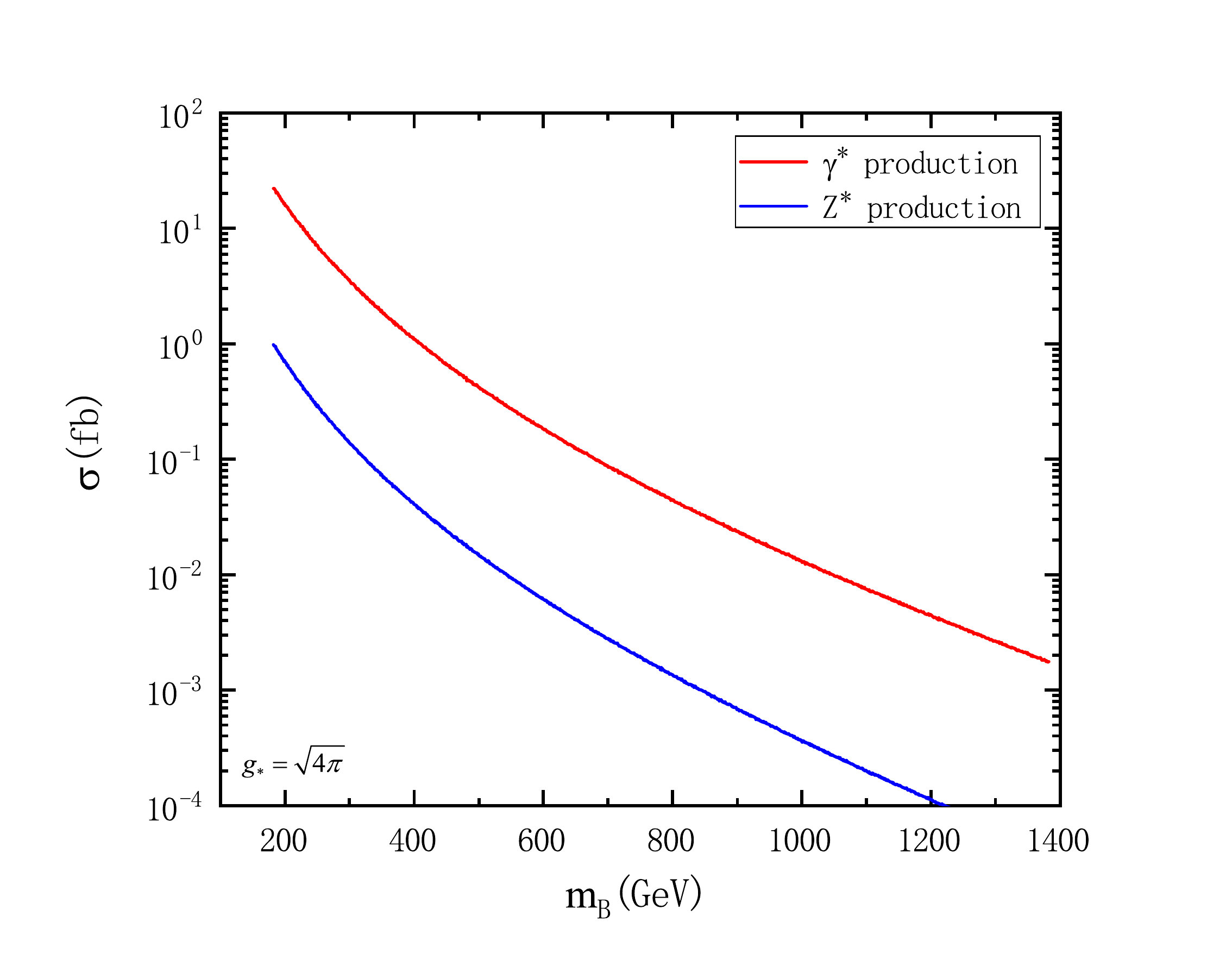}
\centering
\caption{Cross sections of the Drell-Yan processes in Fig.\ref{DY} as function of the lighter mass $m_B$ at the 14-TeV LHC with $g_{*}=\sqrt{4\pi}$, 
as compared to the total cross section of the SM background about $1.13\times10^7$ fb.}
\label{DYc}
\end{figure}

Fig.\ref{DYc} shows the cross sections of the Drell-Yan processes for the lighter composite muon state $\Psi_{B}$ at the  14-TeV LHC with $g_{*}=\sqrt{4\pi}$ that represents a typical value of $g_{*}$ in the parameter space obtained from the muon g-2 anomaly. 
In this figure, the virtual $Z$ contribution is subdominant in comparison with the virtual photon contribution.
This follows from the fact that the coupling of $\Psi_{B}$ pair to $Z$ is relatively smaller than the coupling of $\Psi_{B}$ pair to $\gamma$ as in the SM. 
The SM background for this signal comes from both $pp \rightarrow Z(\ell^{+}\ell^{-})+{\rm jets}$ and $pp \rightarrow \gamma(\ell^{+}\ell^{-})+{\rm jets}$, with leading-order cross section $4.8\times 10^6$ fb and $6.5\times 10^6$ fb respectively. 
Combing them gives a total cross section about $1.13\times10^7$ fb,
implying that the signal cross section as shown in Fig.\ref{DYc} is at least six order smaller than its SM background cross section.

In order to distinguish the above signal from its SM background, 
we apply the latest CMS cuts \cite{CMS:2022cpr, CMS:2022qww} on four-b jets and the CMS cuts \cite{CMS:2015rjz} on
two muons with same flavor but opposite sign
\begin{eqnarray}{\label{cuts}}
{p_{T}}_{j_{1(2,3,4)}} &>& 45 ~{\rm GeV}, \nonumber\\
 {p_{T}}_{\mu_{1(2)}} &>& 20~{\rm GeV} , \nonumber \\
|\eta_{j_{1(2,3,4)}}|&<& 2.5, \nonumber\\
 |\eta_{\mu_{1(2)}}| &<& 2.4,
\end{eqnarray}
to our Drell-Yan processes,
where ${p_{T}}_{j_{1(2,3,4)}}$ and $\eta_{j_{1(2,3,4)}}$ are the transverse momentum
and pseudo-rapidity of jet respectively,
while ${p_{T}}_{\mu_{1(2)}}$ and $\eta_{\mu_{1(2)}}$ are the transverse momentum
and pseudo-rapidity of muon respectively. 
Besides the cuts in Eq.(\ref{cuts}),
any event with an additional jet with $p_T > 30$ GeV or muon with $p_T > 10 $ GeV is rejected.
The cuts in Eq.(\ref{cuts}) can dramatically reduce the number of SM events arising from $pp \rightarrow \gamma(\ell^{+}\ell^{-})+{\rm jets}$,
which makes the SM background of the Drell-Yan processes similar to that of the following $h$-associated processes.

\begin{figure}
\centering
\includegraphics[width=13cm,height=4cm]{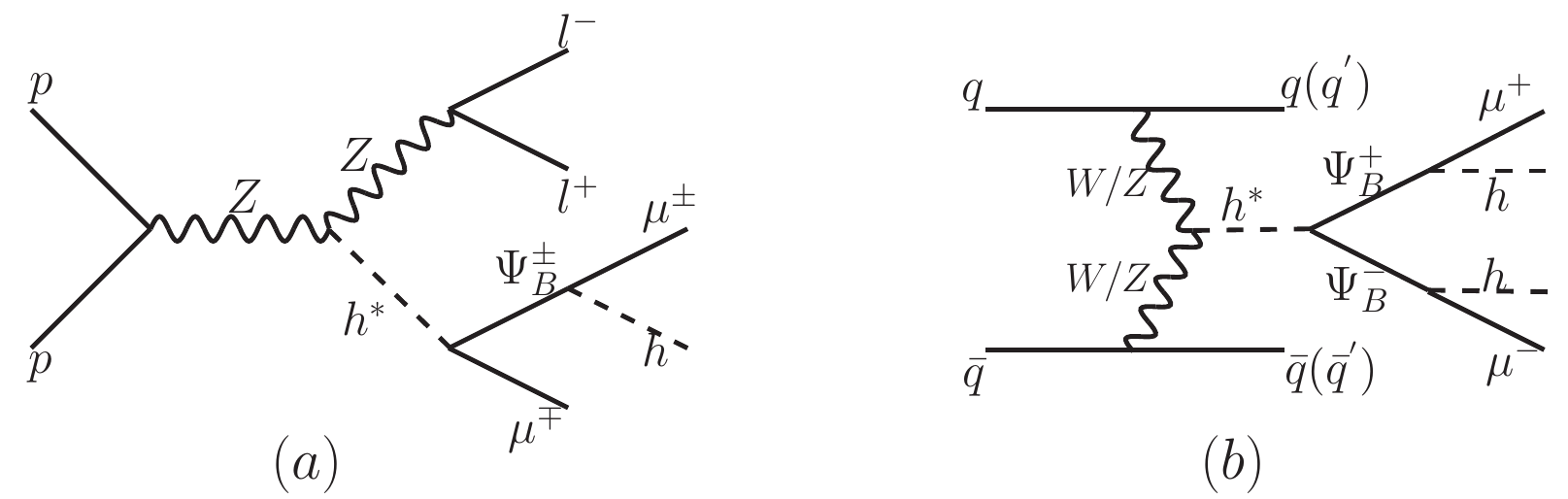}
\centering
\caption{Feynman diagrams for the $h$-associated single (left) and pair (right) production of $\Psi_B$ at the LHC.}
\label{pro}
\end{figure}

\subsubsection{Higgs Portal}
Through Higgs $\Psi_B$ can be singly produced by $pp\rightarrow Zh^{*}\rightarrow Z\Psi^{\pm}_{B}\mu^{\mp}$ which is suppressed by the small mixing angles $\theta_{\mu_{L}}$ and $\theta_{\mu_{R}}$, or produced in pair by $pp\rightarrow \Psi^{+}_{B}\Psi^{-}_{B}jj'$  which is instead not suppressed by the small mixing angles.
Note that the jets $j$ and $j'$ can be either the same or different.
Fig.\ref{pro} shows the Feynman diagrams for these two processes.
In Fig.\ref{pro}(a), 
the single production $pp\rightarrow Zh^{*}\rightarrow l^{+}l^{-}\Psi^{\pm}_{B}\mu^{\mp}\rightarrow l^{+}l^{-}h\mu^{+}\mu^{-}$ is followed by the SM decay $Z\rightarrow l^{+}l^{-}$ with $l=e,\mu$ and the decay mode $\Psi^{\pm}_{B}\rightarrow h\mu^{\pm}$,
for which the SM background is $pp \rightarrow$ mono-Z+jets with cross section of $1.9\times10^{2}$ fb.
In Fig.\ref{pro}(b), the pair production $pp\rightarrow \Psi^{+}_{B}\Psi^{-}_{B}jj'\rightarrow hh\mu^{+}\mu^{-}jj'$ is followed by the decay mode $\Psi^{\pm}_{B}\rightarrow h\mu^{\pm}$,
for which the SM background is $pp \rightarrow Z+$ jets with cross section of $4.8\times10^{6}$ fb. 

\begin{figure}
\centering
\includegraphics[width=13cm,height=8cm]{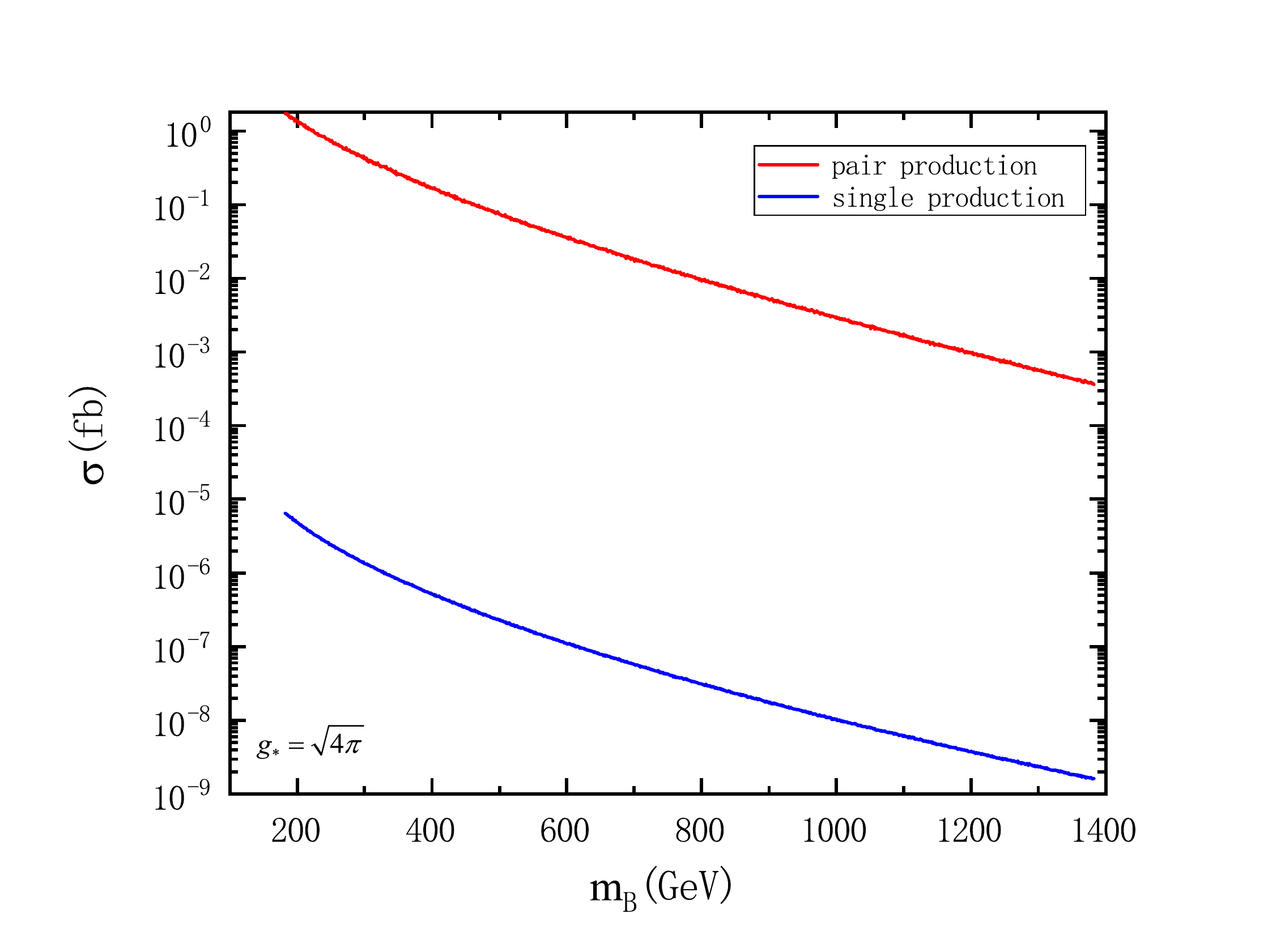}
\centering
\caption{Cross section of the single (in blue) and pair (in red) of $\Psi_B$ in Fig.\ref{pro} as function of $m_B$ at the 14-TeV LHC with $g_{*}=\sqrt{4\pi}$.
Compared to the pair production, 
the cross section of single production is suppressed by the small mixing angles.}
\label{cs}
\end{figure}

Fig.\ref{cs} shows the cross sections of the two processes in Fig.\ref{pro} at the 14-TeV LHC with $g_{*}=\sqrt{4\pi}$.
In the case of single production we have additionally chosen the equal small mixing angles $\theta_{\mu_{L}}\approx\theta_{\mu_{R}}$.
Adjusting the allowed values of these small mixing angles may yield a larger cross section for the single production.
However, it is still not enough to let this process be important.
In contrast, the pair production seems more promising,
as the gap between the cross sections of the signal and SM background is smaller in the mass range $m_B$ less than 300 GeV.
As the final states of this signal contain four b-jets, 
the CMS cuts in the previous Drell-Yan processes can be similarly applied to it.
Then, one recognizes that the SM background of the $h$-associated pair production is similar to that of the Drell-Yan processes.
Given the same $m_B$ and $g_{*}$,
the signal cross section in the $h$-associated pair production in Fig.\ref{cs} is smaller than in the Drell-Yan productions in Fig.\ref{DYc} by roughly an order of magnitude,
leading to the later one more competitive than the former one.
We will neglect the $h$-associated pair production in the following numerical analysis.

\section{Numerical Results}
\label{results}
\begin{figure}
\centering
\includegraphics[width=13cm,height=8cm]{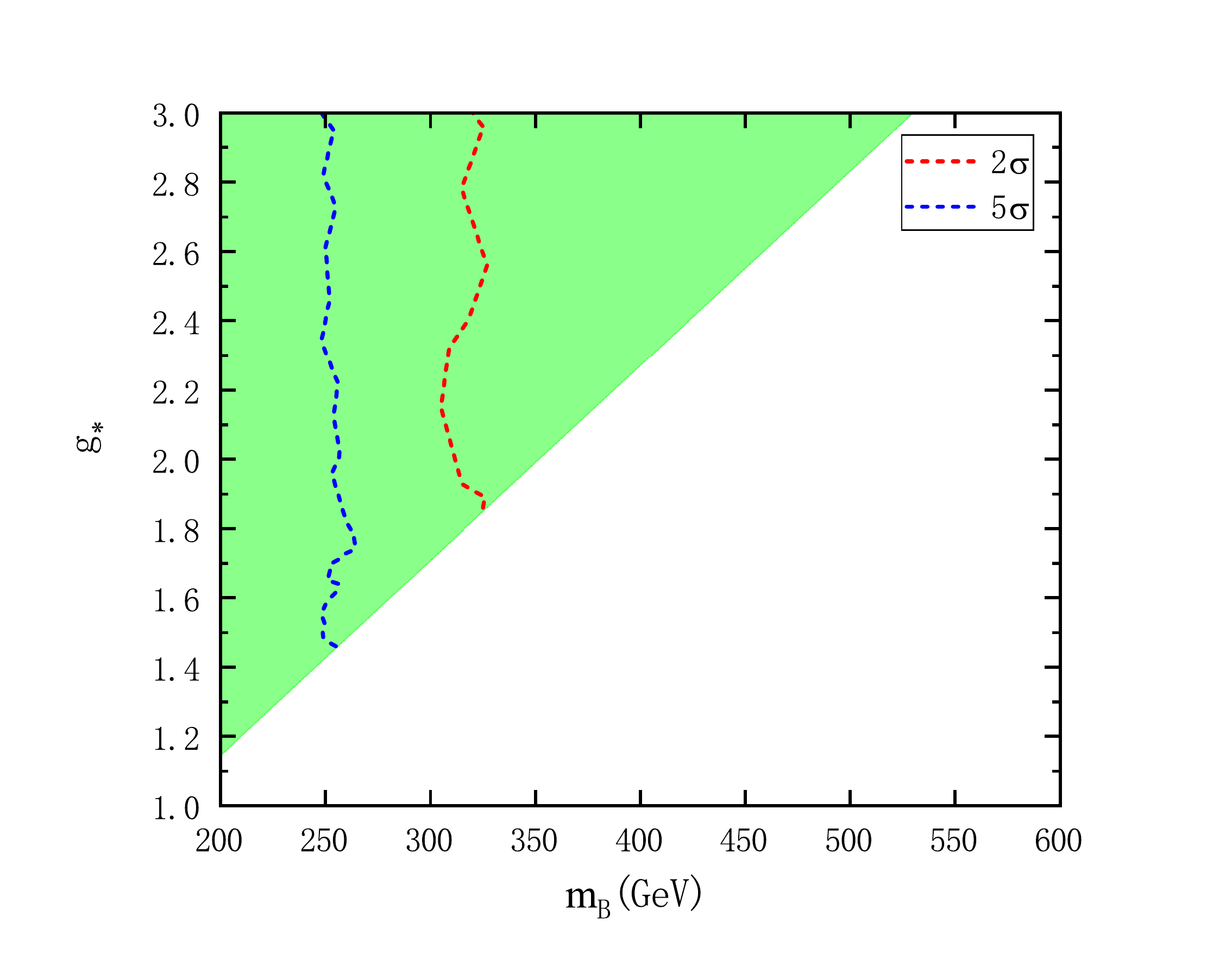}
\centering
\caption{Parameter space (in green) with respect to the muon g-2 anomaly for  $\sin\theta_{\mu_{L}}=\sin\theta_{\mu_{R}}$, 
and the prospect of searching for the parameter regions based on the Drell-Yan processes of $\Psi_B$  at the 14-TeV LHC with an integrated luminosity 3 ab$^{-1}$, 
with 2$\sigma$ exclusion (in red) and 5$\sigma$ discovery (in blue) shown simultaneously. See text for details.}
\label{tot}
\end{figure}

Collecting the parameter space in Sec.\ref{anomaly} and the indirect and direct constraints in Sec.\ref{constraint},
we present the main numerical results in Fig.\ref{tot} for $\sin\theta_{\mu_{L}}=\sin\theta_{\mu_{R}}$.
In this figure, the parameter space (in green band) means the solution to the muon g-2 anomaly within a discrepancy of $2\sigma$, where $g_{*}\sim 1-3$ and $m_{B}\sim 200-550$ GeV.
Unlike light degree(s) of freedom in vectorlike models as required to resolve the anomaly,  
the allowed large mass range of $m_{B}$ manifests the composite Higgs model.
In the light of the direct limits from the Drell-Yan productions of the lighter composite state $\Psi_B$ at the 14-TeV LHC with the integrated luminosity of 3 ab$^{-1}$ with 2$\sigma$ exclusion (in red curve) and 5$\sigma$ discovery (in blue curve)\footnote{We follow the criteria $S/\sqrt{B}=1.96$ for exclusion and $S/\sqrt{S+B}=5$ for discovery, where $S$ and $B$ refer to the event number of signal and SM background, respectively.}, 
the parameter regions with $m_{B}$ up to at most $\sim 255$ GeV can be discovered  and up to $\sim 325$ GeV excluded, respectively. 

A few comments are in order regarding the numerical results in Fig.\ref{tot}.
\begin{itemize}
\item  With $\sin\theta_{\mu_{L}}=\sin\theta_{\mu_{R}}$ all of the indirect constraints in Sec.\ref{indirect} can be satisfied by $f$ larger than $
\sim 3/g_{*}$ TeV.
In particular, the deviation of the decay width of $h\rightarrow \gamma\gamma$ relative to the SM reference value due to the composite muon partners is of order $\sim 10^{-8}\sim 10^{-9}$ in the parameter regions of Fig.\ref{tot}.
In practice, this indirect constraint is more viable for the composite top partners depending on the top partner mass scale.
As the magnitude of the compositeness parameters  grows from $\sim 0.01$ to $\sim 0.1-1$ as seen in Eq.(\ref{Yukawa}), 
the effect of the top partners on this decay is significantly amplified.

\item  The direct limits arising from the Drell-Yan productions are more sensitive to $m_B$ than $g_*$.
The reason is that the signal cross section is affected by $g_*$ only through the dependence of the branching ratio of the decay mode $\Psi^{\pm}_{B}\rightarrow h\mu^{\pm}$ on this parameter,
which is nearly fixed in the parameter regions with $m_{B}\geq 200$ GeV and  $g_{*}\geq 1$. 
We have verified this point by observing that the cross section in Fig.\ref{DYc} hardly change as $g_*$ varies.
\item As noticed in Eq.(\ref{lepton}),
the parameter space is rather sensitive to the mixing effects between the composite muon states. 
If one let the input masses $m_L$ and $m_E$ be highly non-degenerate, 
the parameter space will be obviously reduced or even disappear. 
\end{itemize}

\section{Conclusion}
\label{con}
In this work we have considered the scenario of composite Higgs with partial compositeness as the solution to the muon anomalous magnetic moment as recently confirmed by the FNAL experiment. 
The anomaly is resolved by the one-loop correction to $a_{\mu}$ due to the composite muon partners with their mass scale of order TeVs and strong coupling constant $g_{*}$ of order unity.
Compared with alternative scenarios where light states are required, 
the relatively large mass regions can manifest the composite Higgs model.
We have presented the parameter space by taking into account the constraints of precise tests on $h$, $Z$ and oblique electroweak parameters, 
and explored the prospect of searching for the parameter regions 
in terms of both the Drell-Yan and $h$-associated productions of the lighter composite muon partner at the 14-TeV LHC
with the integrated luminosity 3 ab$^{-1}$.
To draw the direct detection limits, 
we have applied the latest CMS cuts about  four b-jets to both the Drell-Yan and the $h$-associated pair productions.
It turns out that
the parameter regions with the lighter composite muon mass below $\sim 325$ GeV can be excluded at $2\sigma$ order in the Drell-Yan productions, 
despite the ability of exclusion in the $h$-associated productions being weak.
Finally, although not discussed here, this scenario may be linked to other leptonic anomalies.

\section*{Acknowledgments}
This work is supported in part by the National Natural Science Foundation of China (No. 11775039) and the High-level Talents Research and Startup Foundation Projects for Doctors of Zhoukou Normal University (No.ZKNUC2021006).

\end{document}